\documentclass[graybox]{svmult}

\usepackage{mathptmx}       
\usepackage{helvet}         
\usepackage{courier}        
\usepackage{type1cm}        
\usepackage{makeidx}         
\usepackage{graphicx}        
\usepackage{multicol}        
\usepackage[bottom]{footmisc}
\usepackage{xspace}
\usepackage{amssymb}
\usepackage{textcomp}

\newcommand{\tc}{$T_c$\xspace}

\newcommand{\tst}{$T_S$\xspace}
\newcommand{\laf}{LaFeAsO$\mathrm{_{1-x}}$F$\mathrm{_{x}}$\xspace}
\newcommand{\lao}{LaFeAsO\xspace}

\newcommand{\smf}{SmFeAsO$\mathrm{_{1-x}}$F$\mathrm{_{x}}$\xspace}

\newcommand{\bfca}{BaFe$\mathrm{_{2-x}}$Co$\mathrm{_x}$As$\mathrm{_2}$\xspace}
\newcommand{\eufca}{EuFe$\mathrm{_{2-x}}$Co$\mathrm{_x}$As$\mathrm{_2}$\xspace}

\newcommand{\cfca}{CaFe$\mathrm{_{2-x}}$Co$\mathrm{_x}$As$\mathrm{_2}$\xspace}

\newcommand{\tco}{\ensuremath{T_{CO}}\xspace}
\newcommand{\tso}{\ensuremath{T_{SO}}\xspace}
\newcommand{\tlt}{\ensuremath{T_{LT}}\xspace}
\newcommand{\tht}{\ensuremath{T_{HT}}\xspace}

\newcommand{\lasrx}{\ensuremath{\mathrm{La_{2-x}Sr_xCuO_{4}}}\xspace}
\newcommand{\labax}{\ensuremath{\mathrm{La_{2-x}Ba_xCuO_{4}}}\xspace}

\newcommand{\landx}{\ensuremath{\mathrm{La_{1.6-x}Nd_{0.4}Sr_xCuO_{4}}}\xspace}

\newcommand{\lesco}{\ensuremath{\mathrm{La_{1.8-x}Eu_{0.2}Sr_xCuO_4}}\xspace}

\newcommand{\lasrre}{\ensuremath{\mathrm{La_{2-x-y}RE_ySr_xCuO_{4}}}\xspace}

\makeindex             

\begin{document}

\title*{Nernst effect of iron pnictide and cuprate superconductors: signatures of spin density wave and stripe order}
\titlerunning{Nernst effect of iron pnictide and cuprate superconductors}
\author{Christian Hess}
\institute{Christian Hess \at IFW-Dresden, Institute for Solid State Research, P.O. Box 270116, D-01171 Dresden, Germany, \email{c.hess@ifw-dresden.de}}
%
%
\maketitle

\abstract{The Nernst effect has recently proven a sensitive probe for detecting unusual normal state properties of unconventional superconductors. 
In particular, it may sensitively detect Fermi surface reconstructions which are connected to a charge or spin density wave (SDW) ordered state, and even fluctuating forms of such a state.
Here we summarize recent results for the Nernst effect of the iron pnictide superconductor $\rm LaO_{1-x}F_xFeAs$, whose ground state evolves upon doping from an itinerant SDW to a superconducting state, and the cuprate superconductor $\rm La_{1.8-x}Eu_{0.2}Sr_xCuO_4$ which exhibits static stripe order as a ground state competing with the superconductivity.
In \laf, the SDW order leads to a huge Nernst response, which allows to detect even fluctuating SDW precursors at superconducting doping levels where long range SDW order is suppressed. This is in contrast to the impact of stripe order on the normal state Nernst effect in \lesco. Here, though signatures of the stripe order are detectable in the temperature dependence of the Nernst coefficient, its overall temperature dependence is very similar to that of \lasrx, where stripe order is absent. The anomalies which are induced by the stripe order are very subtle and the enhancement of the Nernst response due to static stripe order in \lesco as compared to that of the pseudogap phase in \lasrx, if any, is very small.}

\section{Introduction}
\label{sec:intro}
The Nernst effect is the generation of a transverse electric field $\bf E$ upon the application of a magnetic field $\bf B$ perpendicular to a longitudinal thermal gradient $\nabla T$, i.e., ${\bf E \perp B} \perp \nabla T$. The Nernst \textit{signal} is then defined as the measurable voltage per temperature difference: $e_y=|{\bf E}|/|\nabla T|=E_y/|\nabla T|$, and the Nernst \textit{coefficient} is defined as $\nu=e_y/B$ (see, e.g. \cite{Behnia2009,Wang2001}).
One may relate the Nernst coefficient to other accessible transport quantities through the relation
\cite{Wang2001,Sondheimer1948}
\begin{equation}
{\nu=(\frac{\alpha_{xy}}{\sigma}-S\tan \theta)\frac{1}{B}}. \label{sondheimer}
\end{equation}
Here $S$ is the Seebeck coefficient, $\tan \theta$ the Hall angle, $\sigma$ the electrical conductivity, and $\alpha_{xy}$ the off-diagonal Peltier conductivity. In a one-band metal, the two terms on the right-hand side of Eq.~\ref{sondheimer} cancel exactly if the Hall angle is independent of energy ('Sondheimer cancellation') \cite{Behnia2009,Wang2001,Sondheimer1948}. However, in a multiband electronic structure which may arise from the inherent multi-orbital nature of the electronic states at the Fermi level, or from a Fermi surface reconstruction arising in a charge or spin density wave ordered state, this cancellation is no longer valid. The degree of its violation can be determined experimentally by comparing the measured $\nu$ with the term $S\tan\theta/B$, which can be calculated from electrical resistivity, thermopower, and Hall data.

A little more than ten years ago, the Nernst effect of unconventional superconductors began to attract considerable attention \cite{Behnia2009,Wang2001,Xu2000,Wang2006,Cyr-Choiniere2009,Hackl2009a,Hackl2010,Daou2010,Hackl2009,Fournier1997,Li2007,Kondrat2011,Hess2010}. One reason is that for type-II superconductors it is strongly enhanced by movement of magnetic flux lines (vortices) \cite{Huebener1969,Otter1966,Hagen1990,Ri1994}, where the Nernst coefficient $\nu$ is directly proportional to the drift velocity of the vortices, which has rendered this transport quantity a valuable tool for studying their dynamics. This very fundamental property was used to interpret the unusual enhancement of the Nernst coefficient in the normal state of cuprate high \tc superconductors at temperatures much higher than the critical temperature $T_c$ as the signature of vortex fluctuations \cite{Wang2001,Xu2000,Wang2006}. More specifically, it was proposed that in the pseudogap phase above $T_c$ long-range phase coherence of the superconducting order parameter is lost while the pair amplitude remains finite.
One more recent proposal to explain an unusual Nernst response in the cuprates was that Fermi surface distortions due to stripe or spin density wave (SDW) order could lead to an enhanced Nernst effect \cite{Cyr-Choiniere2009,Hackl2009a,Hackl2010,Hess2010}. In particular, for stripe ordering \lesco and \landx, an enhanced positive Nernst signal at elevated temperature has been associated with a Fermi surface reconstruction due to stripe order \cite{Cyr-Choiniere2009}. Furthermore, a strong anisotropy of the Nernst coefficient arising from the broken rotation symmetry of electron-nematic order has been discussed both experimentally and theoretically \cite{Daou2010,Hackl2009}.

SDW order is also an ubiquitous phenomenon in the second class of high temperature superconductors, the iron pnictide superconductors. However, as compared to the cuprates, much less is known about the Nernst effect of this material class. In a pioneering study Zhu et al. reported an anomalous suppression of the off-diagonal thermoelectric current in optimally doped \laf and suggested the presence of SDW fluctuations near the superconducting transition \cite{Zhu2008}. Matusiak et al. observed a strong enhancement of the Nernst coefficient in the SDW state of the parent compounds and at low doping levels of \cfca and \eufca, but did not find any particular anomaly in the Nernst effect of a purely superconducting doping level that could be attributed to neither vortex flow nor to SDW fluctuations \cite{Matusiak2010, Matusiak2011}. Kondrat et al. systematically investigated the doping-evolution of the Nernst effect in \laf \cite{Kondrat2011}. For the parent compound they observe a huge negative Nernst coefficient accompanied with a severe violation of the Sondheimer cancellation in the SDW state. In their study, a similarly enhanced $\nu$ was observed at underdoped ($x=0.05$) superconducting species, despite the absence of static magnetic order and the presence of bulk superconductivity, strongly suggestive of SDW fluctuations. More conventional transport was observed at optimal doping ($x=0.1$) where the normal state Nernst signal is rather featureless with a more complete Sondheimer cancellation.

The purpose of this paper is to compare the impact of SDW/stripe ordering phenomena on the Nernst coefficient of respective prototype systems of the iron pnictide and cuprate high-temperature superconductors. For the iron pnictides the focus is on the material \laf which up to present appears to represent the rare case that magnetically ordered and superconducting phases are well separated in the electronic phase diagram \cite{Luetkens2009}. For the cuprates the material under scrutiny is \lesco, which is a prototype system exhibiting stripe order over a wide region of the electronic phase diagram \cite{Klauss00}. For each of the considered systems, all considerations and data presented in the next two sections are to a large extent borrowed from two recent studies on \laf and  \lesco by Kondrat et al. \cite{Kondrat2011} and Hess et al. \cite{Hess2010}, respectively.

\section{Nernst effect and SDW fluctuations in the iron-based superconductor $\rm\bf LaFeAsO_{1-x}F_{x}$}
\begin{figure}[t]
\sidecaption[t]
\includegraphics[width=0.8\textwidth]{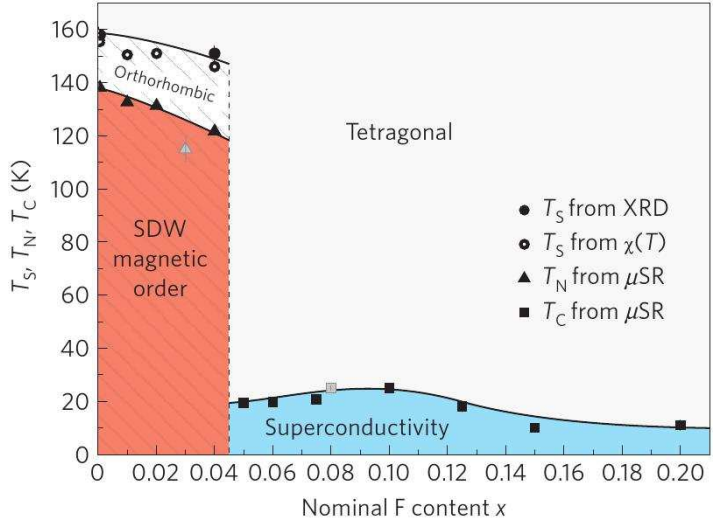}
%
%
\caption{Electronic phase diagram of \laf. The doping dependence of the magnetic and superconducting transition temperatures determined from the \textmu{SR} experiments. Also shown are the tetragonal-to-orthorhombic structural transition temperatures \tst determined directly from X-ray diffraction and from susceptibility measurements. Reproduced from \cite{Luetkens2009}.}
\label{fig:luetkens_phd}       
\end{figure}

In the year 2008, the discovery of superconductivity in \laf \cite{Kamihara2008} initiated a tremendous research effort which yielded soon after a large variety of new superconducting iron pnictide compounds with \tc up to 55~K \cite{Ren2008c}. Figure~\ref{fig:luetkens_phd} reproduces the electronic phase diagram of this compound from reference \cite{Luetkens2009}.
The parent compound \lao is a poor metal and exhibits, as can be inferred from the figure, an antiferromagnetic SDW ground state. The transition towards the SDW state occurs at $T_N=137$~K and is accompanied by a structural tetragonal-to-orthorhombic transition at $T_s\approx160$~K \cite{Luetkens2009,Cruz2008,Klauss2008,Kondrat2009,Qureshi2010}
Upon substituting fluorine for oxygen the SDW phase is destabilized, i.e. $T_s$ and $T_N$ gradually decrease and at some finite doping level ($x\lesssim0.05$) superconductivity emerges. The actual nature of the doping-driven transition from SDW to superconductivity is much under debate. There is evidence that in \laf the transition is abrupt and first order-like towards a homogeneous superconducting state \cite{Luetkens2009} while in other systems (e.g. \smf or \bfca) experiments suggest a finite doping interval where superconductivity and static magnetism coexist \cite{Drew2009,Ni2008}. The obvious proximity to antiferromagnetism suggests spin fluctuations being important for the mechanism of superconductivity with a respective impact on the normal state properties, including the normal state transport \cite{Kondrat2011,Kondrat2009,Prelovsek2009,Prelovsek2010,Hess2009}. 

\begin{figure}[t]
\sidecaption[t]
\includegraphics[scale=.85]{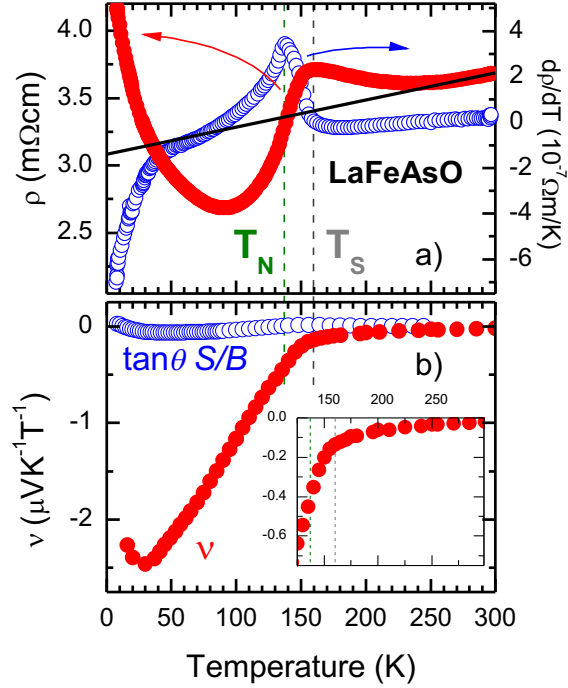}
%
%
\caption{Normalized resistivity $\rho(T)$ and the derivative $d\rho/dT$ (a) and Nernst coefficient $\nu$ (full circles) and $S\tan\theta/B$ (open circles) of \lao as a function of temperature (b). The solid line is a guide to the eye. Data reproduced from \cite{Kondrat2011,Hess2009}.}
\label{fig:undop_rho_nu}       
\end{figure}

The normal state transport properties of the samples from which the phase diagram has been constructed have been studied in great detail \cite{Kondrat2011,Kondrat2009,Hess2009}, where
Kondrat et al. have focused on the Nernst effect of three samples out of this phase diagram for the doping levels $x=0$, 0.05, 0.1 \cite{Kondrat2011}.
Figure \ref{fig:undop_rho_nu} (a) presents the electrical resistivity, $\rho(T)$, of the parent compound \lao as a function of temperature $T$ \cite{Kondrat2009,Hess2009}. $\rho(T)$ develops a deviation from a standard metallic linear $T$-de\-pen\-dence near 300 K upon cooling which leads to a maximum at $T_s$ and a subsequent sharp drop with an inflection point at $T_N$ (visible through an inflection point in $\rho$ and hence a sharp peak in $d\rho/dT$) \cite{Kamihara2008,Klauss2008,Kondrat2009,Hess2009,Mcguire2008}.
A further decrease of temperature leads to a minimum of $\rho(T)$ at $\sim90$~K followed by a strong low-$T$ upturn. 
The origin of this quite anomalous temperature dependence of the resistivity is not entirely clear. Qualitatively, it seems straightforward, however, to rationalize the observed anomalies in terms of enhanced scattering at $T>T_s$,  presumably arising from fluctuations, and, in the SDW state, a reduced carrier density together with a dramatically reduced carrier scattering rate. In particular, the drastic drop of $\rho(T)$ in the SDW state implies a strong enhancement of the carrier relaxation time. The actual nature of the fluctuations which give rise to the enhanced $\rho(T)$ at $T>T_s$ is uncertain. However, there is strong evidence that SDW fluctuations are present and apparently couple to the charge dynamics. Thermal expansion data reveal an extended fluctuation region in the same regime at $T>T_s$, where the resistivity deviates from linearity \cite{Wang2009a}. This is corroborated by the predominantly phononic heat conductivity $\kappa$ of \lao which exhibits a strong dip-like anomaly at $T\approx T_s$ \cite{Kondrat2009,Mcguire2008} which also signals that structural fluctuations are relevant. Due to the tight relationship of the low-temperature orthorhombic distortion with the SDW state, magnetic fluctuations are likely to accompany the structural fluctuations at $T>T_s$.

Figure~\ref{fig:undop_rho_nu}b) shows the temperature dependence of the Nernst coefficient $\nu(T)$ of \lao \cite{Kondrat2011} in direct comparison with the electrical resistivity. $\nu(T)$ is negative over the whole $T$ range. It  decreases moderately from $\nu=-0.02~\mu\rm VK^{-1}T^{-1}$ at 300~K down to $\nu=-0.2~\mu\rm VK^{-1}T^{-1}$ at about 150~K. As can be seen in the figure, at $T\lesssim150$~K, i.e. almost coinciding with the strong drop in the electrical resistivity upon the onset of SDW order, a large negative contribution becomes apparent. The slope of $\nu(T)$ changes strongly and the Nernst coefficient falls towards a large negative value of $-2.5~\mu\rm VK^{-1}T^{-1}$ at around 25~K. Qualitatively, this strong enhancement of the Nernst coefficient should be attributed to the Fermi surface reconstruction that is associated with the SDW phase \cite{Hackl2009a,Kondrat2011}. The value of the Nernst coefficient in the SDW state is remarkably large, because it is about one order of magnitude larger than that generated by vortex flow in the superconducting samples (see below) or in, e.g., cuprate superconductors \cite{Wang2006,Ri1994} which is often considered as a benchmark for a large Nernst effect. Note that a qualitatively similar but quantitatively one order of magnitude smaller impact of SDW order on the Nernst effect has been observed also in CaFe$_2$As$_2$ and EuFe$_2$As$_2$ by Matusiak et al. \cite{Matusiak2010,Matusiak2011}. 

Kondrat et al. have investigated to what extent the 'Sondheimer cancellation' is violated in the SDW phase by comparing the term $S\tan\theta/B$ (which can be easily computed from thermopower, Hall, and resistivity data) \cite{Kondrat2011,Kondrat2009,Hess2009}. The direct comparison of this quantity with the Nernst coefficient reveals clearly $|\nu|\gg |S\tan\theta|/B$, i.e. a \textit{severe} violation of the Sondheimer cancellation in the SDW phase (see Figure~\ref{fig:undop_rho_nu}b)).

\begin{figure}[t]
\sidecaption[t]
\includegraphics[scale=.85]{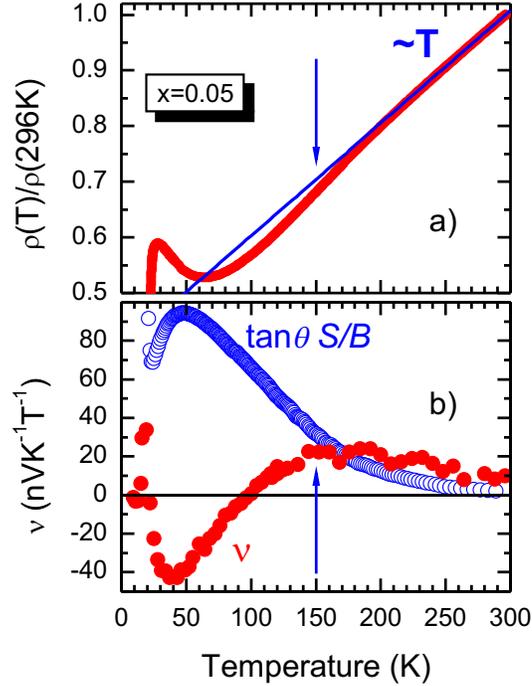}
%
%
\caption{Normalized resistivity $\rho(T)$ (a) and Nernst coefficient $\nu$ (full circles) and $S\tan\theta/B$ (open circles) of \laf at $x=0.05$ as a function of temperature (b). The solid line shows a linear fit to the high-temperature resistivity. Arrows mark the onset of non-linearity in the resistivity and of a strong negative contribution to the Nernst coefficient. Data reproduced from \cite{Hess2009,Kondrat2011}.}
\label{fig:5dop_rho_nu}       
\end{figure}

Superconductivity with rather high critical temperature \tc abruptly emerges in \laf approximately at the doping level $x=0.05$ \cite{Luetkens2009}. The normal state resistivity $\rho(T)$ drastically changes as compared to that of  \laf at $x\leq0.04$ which still exhibit SDW order \cite{Kondrat2009,Hess2009}. A low-$T$ upturn ($T\lesssim 60$~K) is still present before entering the superconducting state at $T_c\approx 21$~K, which is reminiscent of the low-$T$ upturn of the parent compound. At high temperature, however, the clear features at $\sim150$~K of the non superconducting samples have disappeared and $\rho$ increases monotonically for $T\gtrsim60$~K up to 300~K. Hess et al. have pointed out a surprising feature at intermediate temperature \cite{Hess2009}: while $\rho(T)$ becomes linear at $T\gtrsim250$~K, it drops below the low-$T$ extrapolation of this linearity (cf. Figure~\ref{fig:5dop_rho_nu} (a)). Based on the similarity to the SDW-anomalies in the resistivity of the parent compound, it has been suggested that fluctuations connected to the SDW should still be present, despite the suppression of the actual structural and magnetic transitions \cite{Luetkens2009,Qureshi2010}.

The temperature dependence of the Nernst coefficient of this 'underdoped' sample is reproduced in Figure~\ref{fig:5dop_rho_nu}b) \cite{Kondrat2011}. In the superconducting state a strong positive contribution arising from vortex motion is present which extends up to about 40~K. At higher temperature a surprising similarity of $\nu(T)$ with that of the parent compound becomes apparent: Between 300~K and about 150~K, $\nu(T)$ is rather weakly temperature dependent, but at around 150~K the $T$-dependence changes and a sizable negative contribution leads to a sign change at $\sim100$~K and a minimum at $\sim 40$~K where the positive contribution from vortex motion sets in. Kondrat et al. pointed out that in this low-temperature regime (i.e., $T\lesssim150$~K) $|\nu|\approx |S\tan\theta|/B$, i.e. a significant violation of the Sondheimer cancellation is still present. Furthermore, the negative contribution between 40~K and 150~K, despite a strongly reduced magnitude as compared to that of the parent compound is still of similar size as the vortex contribution at low $T$ \cite{Kondrat2011}. 

The strong similarity of the anomaly at $T\lesssim150$~K with the SDW-enhanced Nernst coefficient of the parent compound suggests that SDW order should be also considered in this superconducting sample. However, as already mentioned above, the material exhibits bulk superconductivity, whereas muon spin relaxation (\textmu{SR}) and M\"ossbauer spectroscopy show no trace of magnetic ordering in this $T$-regime \cite{Luetkens2009,Qureshi2010}. It seems therefore straightforward to conclude that an SDW precursor, in the form of fluctuations or possibly nematic phases give rise to the enhanced Nernst response \cite{Hackl2009a,Hackl2009,Kondrat2011}. Thus the afore notion, based on the electrical resistivity that SDW fluctuations are present in the normal state of underdoped, superconducting \laf \cite{Hess2009} is strongly corroborated. This is consistent with the observation for the parent compound \cite{Kondrat2011} that at $T\gtrsim T_N$, i.e. in a $T$-range where SDW precursors are truly present \cite{Wang2009a,Nakai2008}, the Nernst response is enhanced with a similar magnitude as in the low-$T$ regime of the underdoped material (c.f. inset of Figure~\ref{fig:undop_rho_nu}b)). Note, that the observed negative sign of the SDW-related contribution to the Nernst coefficient unambiguously rules out vortex fluctuations \cite{Wang2001,Xu2000,Wang2006} as a thinkable origin since these should give rise to a positive Nernst response.

\begin{figure}[t]
\sidecaption[t]
\includegraphics[scale=.85]{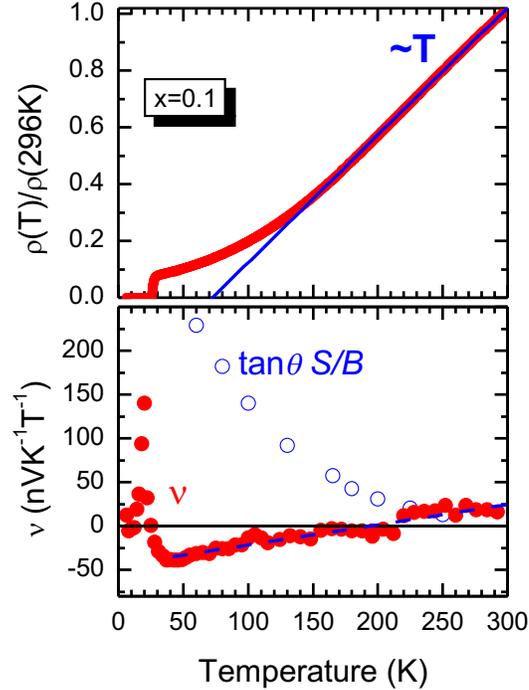}
%
%
\caption{Normalized resistivity $\rho(T)$ (a) and Nernst coefficient $\nu$ (full circles) and $S\tan\theta/B$ (open circles) of \laf at $x=0.1$ as a function of temperature (b). The solid line shows a linear fit to the high-temperature resistivity.  The dashed line  is a guide to the eye. Data reproduced from \cite{Hess2009,Kondrat2011}.}
\label{fig:10dop_rho_nu}       
\end{figure}

The enhancement of the doping level to optimal doping $x=0.1$ leads to drastic changes of both the resistivity and the Nernst effect: in the resistivity, instead of a low-$T$ upturn above \tc, a quadratic increase is observed up to  $\sim150$~K, i.e. $\rho(T)=\rho_0+AT^2$ ($\rho_0=\mathrm{const}$). The resistivity drop at $\sim150$~K has practically disappeared and a smooth crossover to a linear high-$T$ behavior is present (cf. Figure~\ref{fig:10dop_rho_nu}a)) \cite{Hess2009}.

In the Nernst response, despite a very similar behavior in the vicinity of \tc and similar magnitude as compared to the underdoped compound a completely different normal state behavior is observed. In the whole normal state at $T\gtrsim40$~K, $\nu(T)$ is featureless with a weak positive slope. In particular, no anomaly similar to that of the underdoped material is present. The Sondheimer cancellation is more complete now, i.e. $|\nu|\ll |S\tan\theta|/B$ is found at low $T$. Both, the Fermi liquid-like resistivity and rather conventional Nernst effect suggests that \laf at optimal doping displays more normal metallic properties as compared to those of the underdoped and undoped levels. In particular, all features, which  could be related to SDW order and or SDW fluctuations are absent.

The doping dependence of the SDW signature in the Nernst response suggests that the material \laf evolves from a very unusual metal at $x=0$ to a more conventional one at $x=0.1$, where at $x=0.05$ the interesting situation of a fluctuating/nematic SDW state appears to be realized. Here, the Nernst effect turns out as a very sensitive probe to this quite subtle state which is not detectable by diffraction techniques or local probes such as \textmu{SR} or M\"o{\ss}bauer spectroscopy \cite{Luetkens2009,Qureshi2010}.

\section{Nernst effect and stripe order in the cuprate superconductor $\rm\bf La_{1.8-x}Eu_{0.2}Sr_xCuO_4$}
In cuprate superconductors, the tendency towards the segregation of spins and holes is much under debate with respect to the nature of superconductivity and the pseudogap phase \cite{Cyr-Choiniere2009,Daou2010,Tranquada95,Kivelson98,Tranquada2004,Hinkov2004,Hinkov2007,Kohsaka2007,Hinkov2008}. 
Clear evidence for \textit{static stripe order} has been observed in materials which are closely related to the fundamental cuprate superconducting system \lasrx. Prototype materials exhibiting static stripes are the compounds \labax \cite{Tranquada2004,Fujita2004,Hucker2010,Tranquada2008,Wen2008a,Xu2007} and the closely related \lesco and \landx \cite{Tranquada95,Fink2009a,Fink2011}. In the case of stripe order, these materials exhibit stripe-like arrangements of alternating hole-rich and antiferromagnetic regions, where in all these materials an intimate interplay between structure, stripe order and superconductivity is present. More specifically, bulk superconductivity is suppressed in favor of static stripe order where the latter is stabilized through a particular tilting pattern of the $\rm CuO_6$ octahedra in the low-temperature tetragonal structural phase (LTT-phase) \cite{Klauss00,Tranquada95,Tranquada2008,Buechner1991,Buechner1993,Buchner94}. 

In the prototype stripe ordering compound \labax the LTT phase is only present in a limited doping range around $x=1/8$ \cite{Axe1989}. At this very doping level, the LTT phase and therefore static stripe order is present only at relatively low temperature $T\lesssim55$~K, where the stripe order sets in abruptly directly at the transition to the LTT phase \cite{Hucker2011,Wilkins2011}. Recently, very intriguing results for the Nernst effect of this materials have been reported \cite{Li2011}, which point to time-reversal symmetry breaking due to the stripe order. The reported onset of a spontaneous Nernst signal related to the stripe order deserves further attention which is, however, out of the scope of this overview. 

\begin{figure}[t]
\sidecaption[t]
\includegraphics[width=0.6\textwidth]{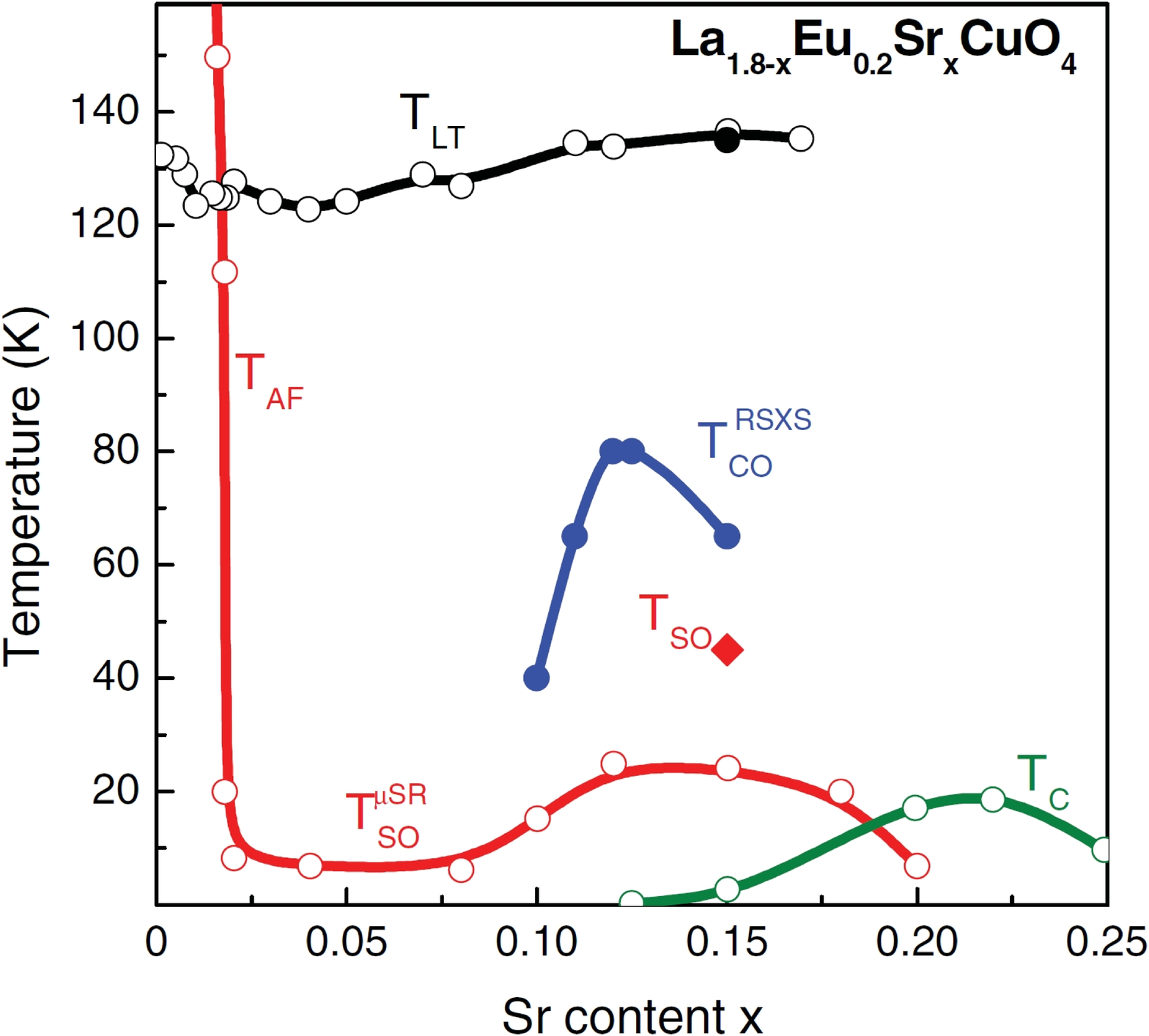}
%
%
\caption{Phase diagram of \lesco showing transition temperatures for the LTT phase \tlt, the antiferromagnetic structure $T_\mathrm{AF}$, the magnetic stripe order \tso, the stripe-like charge order \tco, and the superconducting transition temperature \tc. Closed circles from resonant soft x-ray scattering (RSXS) experiments \cite{Fink2011}. Open circles from Reference \cite{Klauss00}. Closed diamond from neutron diffraction data presented in Ref. \cite{Hucker2007}. Reproduced from \cite{Fink2011}.}
\label{fig:LESCO-phd}       
\end{figure}

Concerning the stabilization of the LTT phase, \lesco is very different as compared to \labax and also \landx. The LTT phase is present at lowest temperature over a wide doping range, see Figure~\ref{fig:LESCO-phd}.  In addition, irrespective of doping, the transition temperature extends up to rather high temperatures $T_{LT}\approx 120\pm10$~K, i.e. much higher than in in \labax and  \landx where $T_{LT}\approx 55$~K and $T_{LT}\approx 70$~K, respectively \cite{Buechner1991,Buechner1993,Hucker2011,Wilkins2011,Tranquada96a}. Bulk superconductivity with a considerable critical temperature \tc  is suppressed in \lesco over a wide doping range up to $x\lesssim0.2$. Around $x=0.2$  the tilt angle of the octahedra and hence the buckling of the plane which decreases with increasing hole doping becomes smaller than a critical value \cite{Klauss00,Buchner94}. At $T>T_{LT}$ the structure enters the low temperature orthorhombic (LTO) phase in which the buckling pattern of the $\rm CuO_2$ planes does not support static stripe order \cite{Klauss00}. At even higher temperatures the structure enters a further tetragonal phase (so-called high temperature tetragonal phase, HTT) at $T_{HT}$. In the case of \lesco, $T_{HT}>300$~K, at $x\leq0.15$ and $T_{HT}\approx220$~K for $x=0.2$ \cite{Klauss00}.

\begin{figure}[t]
\includegraphics[clip,width=0.97\columnwidth]{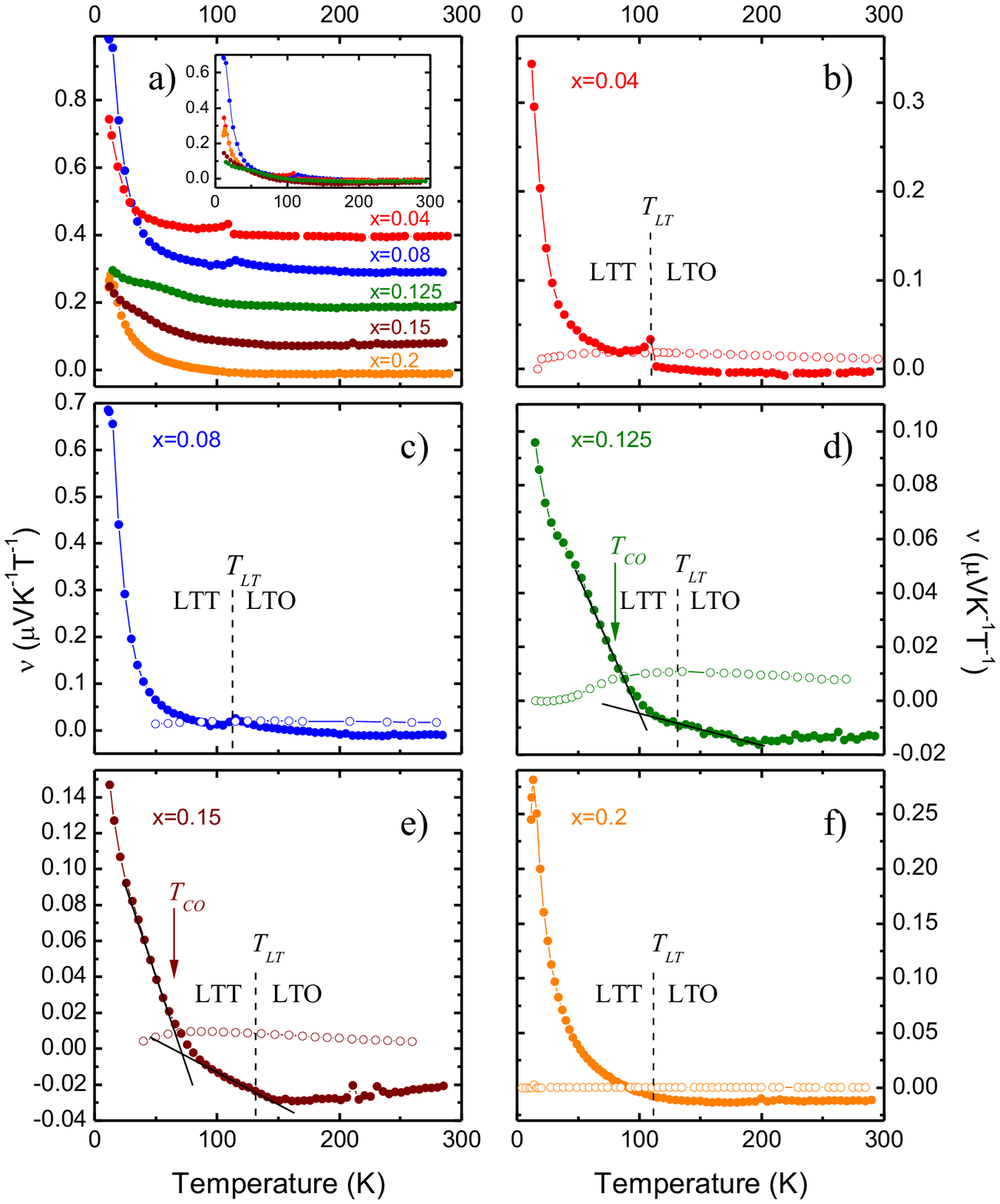}
\caption{Nernst coefficient $\nu$ of \lesco ($x=0.04$, 0.08, 0.125, 0.15, 0.2) as a function of temperature. a) Overview on all data. The presented curves have been shifted for clarity. Inset: all curves at the same scale. b)-f) Temperature dependence of $\nu$ (full symbols) and of $S\tan\theta /B$ (open symbols) for each doping level. Solid lines are linear extrapolations of $\nu(T)$ in order to extract $T_{\nu*}$. Arrows mark the charge stripe ordering temperatures \tco for $x=0.125$, 0.15 as seen in RSXS experiments \cite{Fink2009a}, see Figure~\ref{fig:LESCO-phd}. Reproduced from \cite{Hess2010}.} 
\label{Fig:all_nernst}
\end{figure}

Hess et al. have recently reported the transport properties of \lesco ($x=0.04$, 0.08, 0.125, 0.15, 0.2) with a special focus on the Nernst effect in the normal state \cite{Hess2010}. Figure~\ref{Fig:all_nernst} presents the temperature dependence of the Nernst coefficient $\nu$ of all investigated samples. 

Panel a) of the figure shows an overview on all data. Clearly, the overall magnitude of the Nernst coefficient is very similar in the temperature range which corresponds to the normal state, i.e. at $T\gtrsim50$~K. Relatively small anomalies are present in $\nu(T)$ which will be discussed in more detail further below. At lower temperatures ($T\lesssim50$) all curves strongly increase with falling temperature. However, the magnitude of the increase is quite non-monotonic as a function of doping with a clear minimum at $x=1/8$. This low-temperature rise in the Nernst coefficient can thus be attributed to fluctuations of the superconducting order parameter which experiences a severe suppression in the presence of stripe order \cite{Klauss00}, which is strongest at $x=1/8$.

The individual curves for the Nernst coefficient  $\nu$ (full symbols) of each doping level are separately shown in panels b) to f) of Figure~\ref{Fig:all_nernst}. As was done above for \laf, the quantity $S\tan\theta /B$ (open symbols) is displayed as well \cite{Hess2010} which allows to judge whether any of the observed anomalies is related to  $\alpha_{xy}$, i.e. a true off-diagonal thermoelectric quantity or to an anomalous behavior in the complementary transport coefficients.

Before discussing the potential effect of stripe order on the Nernst response it is interesting to examine the data with respect to any impact of the structural transition at \tlt which is present in all compounds \cite{Klauss00,Hess03a}. Indeed, a jump-like anomaly is present at \tlt for $x=0.04$ and $x=0.08$, where the jump size for the latter is smaller. No anomaly is present at higher doping levels. Interestingly, $S\tan\theta /B$ does not contribute significantly to the observed jumps. This means that the Nernst response, more specifically, $\alpha_{xy}$, directly couples to structural distortions of the $\rm CuO_2$-plane. There is an apparent correlation of the jump size to the degree of buckling of the CuO$_2$ plane, since concomitantly to the decrease of the jump size (towards its complete disappearance at $x\geq0.125$) with increasing Sr doping level the tilting angle of the $\rm CuO_6$ octahedra decreases as well \cite{Buchner94}. However, it seems reasonable that not only structural (degree of buckling) but also electronic details (hole content) play a decisive role in this regard because the anomaly at \tlt decreases very rapidly with increasing doping.

The closer inspection of the data shown in Figure~\ref{Fig:all_nernst} reveals further interesting features which are clearly discernible at $x=0.125$ and $x=0.15$. For $x=0.125$, two kink-like features are present which deserve closer consideration (see figure~\ref{Fig:all_nernst}d). One is is located deep in the LTT phase at $T_{\nu*}\approx100$~K and is connected with a strong change of slope, the other occurs at much higher temperature $T_\nu\approx180$~K, which is in the LTO phase.
The measured onset temperatures of the charge stripe and the spin stripe order which are known as $T_{CO}=80$~K \cite{Fink2009a,Fink2011} and $T_{SO}\approx45$~K \cite{Klauss00,Hucker2007}, respectively. These are clearly at much lower temperature than both $T_{\nu*}$ and $T_{\nu}$, and thus a connection with the observed kinks is not obvious. However, the lower-temperature kink at $T_{\nu*}$ and the charge stripe ordering temperature \tco, detected by RSXS experiments, occur at a not too  different temperature, which brings to mind a possible close connection between both. As all diffraction experiments also the RSXS requires a certain correlation length of the stripe order to be exceeded in order to generate a superlattice reflection. Short range stripe order might already develop at $T_{\nu*}$, giving rise to an enhanced Nernst coefficient at this temperature, but are beyond resolution in RSXS. Only at the onset of long range order at \tco RSXS is able to detect the stripe order. 

On the other hand, the kink-temperature $T_{\nu}$ seems to be too high to account for the stripe ordering phenomena in the LTT phase in an obvious manner. A possible interpretation for the high-temperature anomaly has been suggested by Cyr-Choini\`{e}re et al. who speculated that the high temperature kink could mark the onset of stripe fluctuation which could cause a Fermi surface reconstruction \cite{Cyr-Choiniere2009}. On the other hand, Hess et al. \cite{Hess2010} pointed out that one cannot exclude that subtle structural effects unrelated to electronic order are the actual cause of the slight enhancement at $T_\nu$. For example, soft phonon type precursors of the $\rm LTO\rightarrow LTT$ transition are known to be ubiquitous in the LTO phase of both \lasrre (RE=Rare Earth), which undergoes the $\rm LTO\rightarrow LTT$ transition, and \lasrx, which remains in the LTO phase down to lowest temperature \cite{Pintschovius90,Martinez91,Keimer93}. In fact, this conjecture is supported by the jump-like response of the Nernst coefficient at \tlt for the lower doping levels. It seems noteworthy to mention that Cyr-Choini\`{e}re et al. reported very similar data for \lesco at $x=0.125$ as compared to the data shown in Figure~\ref{Fig:all_nernst}. However, their data exhibit a significantly lower $T_\nu\approx140$~K (which is very close to the temperature of the structural phase transition \tlt and lack a kink at $T_{\nu*}$ \cite{Cyr-Choiniere2009}.

The situation at $x=0.15$ is very similar to that of $x=0.125$. There, $T_{\nu*}\approx70$~K and $T_{\nu}\approx145$~K with respect to $T_{CO}=65$~K \cite{Fink2009a,Fink2011} and $T_{SO}\approx45$~K \cite{Klauss00,Hucker2007}. Note that $T_{\nu*}\approx T_{CO}$ in this case which corroborates the above conjecture that the kink at $T_{\nu*}$ could be related to the formation of static charge stripe order.

\begin{figure}[t]
\includegraphics[clip,width=0.97\columnwidth]{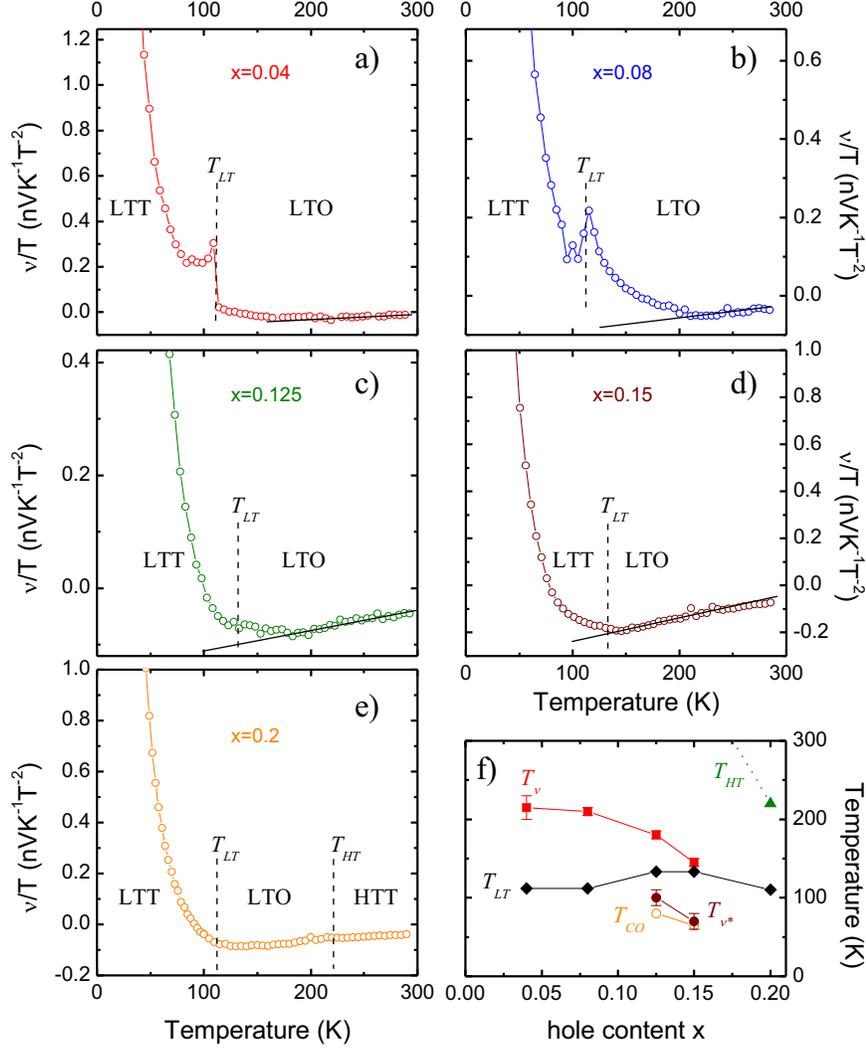}
\caption{a)-e) $\nu/T$ of \lesco ($x=0.04$, 0.08, 0.125, 0.15, 0.2) as a function of temperature. Solid lines are linear extrapolations of the high temperature linear behavior of $\nu(T)$ in order to extract $T_{\nu}$. f) Phase diagram showing \tlt ($\blacklozenge$), \tht ($\blacktriangle$), $T_{\nu}$ ($\blacksquare$) $T_{\nu*}$ (full circles) and \tco (open circles) from RSXS measurements \cite{Fink2009a,Fink2011}. Reproduced from \cite{Hess2010}.} 
\label{Fig_nut}
\end{figure}

Similarly clear anomalies as those seen for $x=0.125$ and $x=0.15$ are not discernible at other doping levels. In order to detect a potential anomalous enhancement of the Nernst coefficient due to  stripe order, Cyr-Choini\`{e}re et al. have suggested to plot the quantity $\nu/T$ versus temperature. This representation relies on the assumption that ordinary normal state quasiparticles should cause a Nernst response which is linear in temperature, when no Fermi surface reconstruction and no contribution from superconductivity are present \cite{Cyr-Choiniere2009}.

Panels a) to e) of Figure~\ref{Fig_nut} display this representation for the afore discussed Nernst effect data of \lesco \cite{Hess2010}. Indeed, at all doping levels up to $x=0.15$, $\nu/T$ is linear at high $T$ and deviates from this linearity at a characteristic temperature. This characteristic temperature decreases monotonically upon increasing doping, where for $x=0.125$ and 0.15 it is identical to that of the high-temperature kink at $T_{\nu}$. One should note that the doping level $x=0.2$, despite  $\nu/T$ being also linear in $T$ at high temperature, should not be considered in this way since this sample undergoes two structural transitions in the region of interest. One is the  $\rm LTO\rightarrow LTT$ transition at $T_{LT}\approx110$~K, the other is the  $\rm HTT\rightarrow LTO$ transition at $T_{HT}\approx220$~K.

Hess et al. have summarized their findings \cite{Hess2010} in the phase diagram shown in Figure~\ref{Fig_nut}f). The main finding from the investigated data on \lesco is the rather good agreement between the lower kink temperature $T_{\nu*}$ and the charge stripe ordering temperature \tco for the doping levels $x=0.125$ and 0.15, where the stripe order has been experimentally detected by diffraction experiments. Qualitatively, the enhanced $\nu$ at $T<T_{CO}$ seems consistent with theoretical results by Hackl et al., who calculated the temperature dependence of the quasiparticle Nernst effect for $p=1/8$ stripe order in a mean field approach \cite{Hackl2010}.

\begin{figure}[t]
\includegraphics[clip,width=1\columnwidth]{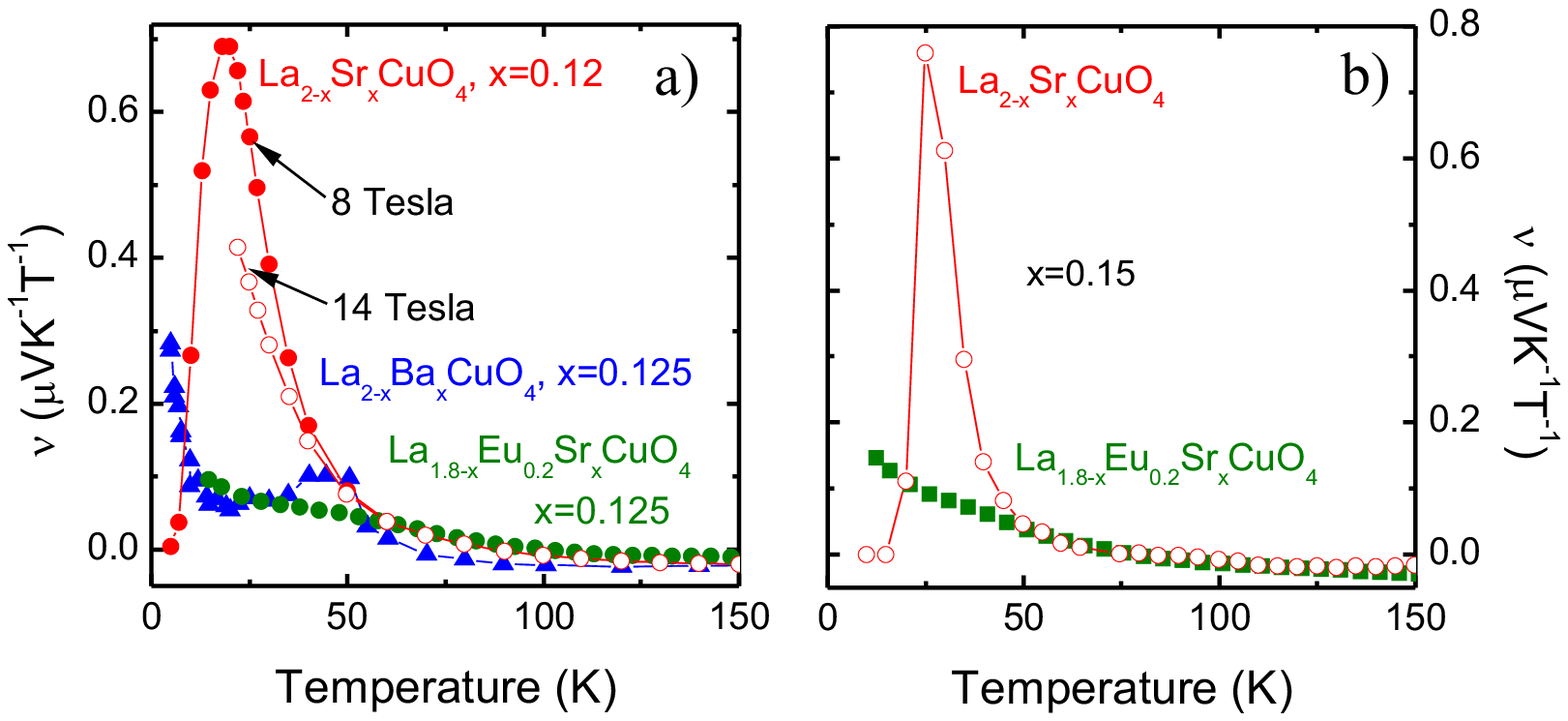}
\caption{Nernst coefficient $\nu$ of \lasrx (circles), \lesco (squares) and \labax (triangles) as a function of temperature. a) \lesco and \labax at $x=0.125$ and \lasrx at $0.12$. b) \lesco and \lasrx at $x=0.15$. Data taken from \cite{Wang2006,Hess2010,Li2011}.} 
\label{Fig_LESCO_LSCO}
\end{figure}

It was further pointed out \cite{Hess2010} that the  salient, stripe order-induced features are, in fact, very subtle anomalies. To illustrate this, they compared the Nernst coefficient of both stripe ordering \lesco and non-stripe ordering, bulk superconducting \lasrx for the doping levels $x\approx0.125$ and $x=0.15$ (see Figure~\ref{Fig_LESCO_LSCO}). Apparently the Nernst-effect of both variants is very similar in the normal state ($T\gtrsim60$~K), in particular, in the vicinity  of the kink anomalies at $T_{\nu}$ and $T_{\nu*}$ in the chosen scale, no significant difference is detectable. Strong differences occur only below $T\approx 60$~K where apparently true superconducting fluctuations lead to a strong enhancement of $\nu$ in \lasrx whereas $\nu$ of \lesco remains much smaller. It is enlightening to compare the findings at $x\approx0.125$ with the very recent data for \labax at this doping level \cite{Li2011} (see Figure~\ref{Fig_LESCO_LSCO}, left panel). As can be seen in the figure, at $T\gtrsim60$~K the data for \labax are very similar to those of \lasrx and \lesco. In the small interval at $T_\mathrm{LT}\approx55~{\rm K}\lesssim T\lesssim60$~K, i.e., as long as the compound is in the LTO phase and no static stripe order is present, the $\nu(T)$ values are very similar to those of \lasrx and become larger than those of \lesco. At $T\lesssim T_\mathrm{LT}$, where static stripe order is present, however, the curve drops strongly and becomes very similar to that of \lesco, which is in the stripe-ordered phase in the entire temperature range below $T_{CO}=80$~K \cite{Fink2009a,Fink2011}.

These observations provide evidence that the magnitude of the Nernst response for static and fluctuating stripes is practically the same (apart from the subtle anomalies at $T_\nu$). Theoretical treatments for the Nernst response in the presence of fluctuating stripes are therefore required. On the other hand, the vortex fluctuation scenario attributes the normal state Nernst coefficient in \lasrx and \labax at $T\lesssim120$~K largely to superconducting fluctuations \cite{Wang2001,Xu2000,Wang2006,Li2011}. In particular, for $x=0.12$ and $x=0.15$ the onset temperatures of such fluctuations have been inferred from a weak increase of $\nu(T)$ at $T\lesssim110$~K and $T\lesssim100$~K, respectively \cite{Wang2001,Xu2000,Wang2006,Li2011}. In view of the similarity between the Nernst effect data in this high-temperature regime for non-stripe ordered \lasrx and \labax (at least at $T>T_{LT}$) and stripe ordered \lesco, a theoretical treatment which is based on vortex fluctuation-enhanced Nernst response should explain why in the presence of stripe order, which suppresses bulk superconductivity, the normal state Nernst effect is practically not affected by the stripe order.

\section{Conclusion}
The above Nernst effect data for the unconventional superconducting systems \laf and \lasrx clearly demonstrate that SDW order (if stripe order of the cuprates is understood as such) has very different impact on this transport quantity, depending on the system. The onset SDW order has a huge effect on the Nernst coefficient of the itinerant antiferromagnet \lao  whereas the effect is tiny in the stripe ordering cuprates \cite{Cyr-Choiniere2009,Daou2010}.
Furthermore, the degree of order has a very different consequence. In \laf at $x=0.05$ the fluctuating SDW signal is about two orders of magnitude smaller than that of the parent compound where static long range order is present. Despite of this strong suppression, the SDW signal is still substantial and exceeds all effects observed in the cuprates and even typical vortex flow signals \cite{Wang2006,Cyr-Choiniere2009,Daou2010,Hess2010,Li2011}. In the cuprates, as was mentioned above, the difference between static and fluctuating stripes is apparently very tiny as can be inferred from the very similar Nernst response of stripe ordered \lesco and \lasrx for which stripes can be presumed to be fluctuating.

\begin{acknowledgement}
Support by the Deutsche Forschungsgemeinschaft through the Research Unit FOR538 (Grant No. BU887/4) and the Priority Programme SPP1458 (Grant No. GR3330/2) is gratefully acknowledged. This work would not have been possible without contributions by U. Ammerahl, G. Behr, B. B\"uchner, A. Revcolevschi, and, in particular, A. Kondrat and E. Ahmed. Furthermore, the author thanks D. Bombor and F. Steckel for proofreading the manuscript.

\end{acknowledgement}

%

\end{document}